\documentclass[10pt,twocolumn,twoside]{IEEEtran}

\usepackage{amsfonts}
\usepackage{amssymb}
\usepackage{cite}
\usepackage[cmex10]{amsmath}
\usepackage[]{algorithm}
\usepackage{algorithmic}
\usepackage{subeqnarray}
\usepackage{cases}
\usepackage{graphicx}
\usepackage{subfigure}
\usepackage{booktabs}
\usepackage{xcolor}
\usepackage{soul}
\usepackage{tabu}
\usepackage{amsmath}
\usepackage{multirow}
\usepackage{threeparttable}
\usepackage{stmaryrd}
\usepackage{makecell}
\usepackage{setspace}
\usepackage{lipsum}
\usepackage{mathtools, cuted}

\usepackage{textcomp}
\usepackage{xcolor}

\newtheorem{theo}{Theorem}

\ifCLASSINFOpdf
\else
\fi

\begin{document}
\title{Deep-Learning-Aided Alternating Least Squares for Tensor CP Decomposition and Its Application to Massive MIMO Channel Estimation}
%
%
%

\author{Xiao Gong,
Wei~Chen,~\IEEEmembership{Senior Member,~IEEE}, Bo~Ai,~\IEEEmembership{Fellow,~IEEE},
Geert Leus,~\IEEEmembership{Fellow,~IEEE}

\thanks{Xiao Gong, Wei Chen and Bo Ai are with State Key Laboratory of Advanced Rail Autonomous Operation, School of Electronic and Information Engineering, Beijing Jiaotong University, Beijing 100044, China (email: \{xiaogong,weich,boai\}@bjtu.edu.cn).}
\thanks{Geert Leus is with the Department of Microelectronics, Delft University of Technology, 2628 CD Delft, Netherlands (e-mail: g.j.t.leus@tudelft.nl).}
}

\maketitle
\begin{abstract}
CANDECOMP/PARAFAC (CP) decomposition is the mostly used model to formulate the received tensor signal in a massive MIMO system, as the receiver generally sums the components from different paths or users. To achieve accurate and low-latency channel estimation, good and fast CP decomposition (CPD) algorithms are desired. The CP alternating least squares (CPALS) is the workhorse algorithm for calculating the CPD. However, its performance depends on the initializations, and good starting values can lead to more efficient solutions. Existing initialization strategies are decoupled from the CPALS and are not necessarily favorable for solving the CPD. This paper proposes a deep-learning-aided CPALS (DL-CPALS) method that uses a deep neural network (DNN) to generate favorable initializations. The proposed DL-CPALS integrates the DNN and CPALS to a model-based deep learning paradigm, where it trains the DNN to generate an initialization that facilitates fast and accurate CPD. Moreover, benefiting from the CP low-rankness, the proposed method is trained using noisy data and does not require paired clean data. The proposed DL-CPALS is applied to millimeter wave MIMO-OFDM channel estimation. Experimental results demonstrate the significant improvements of the proposed method in terms of both speed and accuracy for CPD and channel estimation.
\end{abstract}


\IEEEpeerreviewmaketitle

\section{Introduction}
To reduce the degrading effects in harmful propagation environments, modern wireless communication systems tend to add more degrees of freedom by transmitting signals covering multiple domains, e.g., space, frequency, polarization and/or code, which extends the traditional vector or matrix signals to multidimensional arrays, i.e., tensors. The multidimensional signal structures can be naturally characterized using tensor decomposition models.
In the past decades, tensor models have found a wide range of applications for wireless communication systems \cite{de2016overview}. For example, the CANDECOMP/PARAFAC (CP) decomposition, also called canonical polyadic decomposition \cite{hitchcock1927expression, kolda2009tensor}, which factorizes a tensor into a sum of rank-one tensors, is widely applied to the received signal model \cite{chen2021tensor}, as the receiver generally sums the signal components of different paths or users.

\par
As a remarkable 5G technology, massive multiple-input multiple-output (MIMO) antenna systems have been applied in practice for increasing the data rate and reliability \cite{larsson2014massive}. The large-scale antenna arrays lead to high-dimensional channels, especially when combined with other transmitting domains, such as the frequency domain in orthogonal frequency division multiplexing (OFDM) systems. Channel estimation is crucial to embrace the potential gains of massive MIMO systems \cite{barhumi2003optimal}. However, the acquisition of channel state information (CSI) becomes challenging and expensive in terms of accuracy, training overhead and computational complexity. Fortunately, when adopting the millimeter-wave (mmWave) band or antennas at a high altitude, the sparse scattering
property enables an approximation of the channel using a low-rank model \cite{7727995}. By exploiting this low-rankness of the channel tensor, one can decompose the high-dimensional channels into low-dimensional factors to reduce the storage and enhance the estimation of the channel or its latent parameters, e.g., angles of arrival and departure (AoAs/AoDs), delays, Doppler shifts and path gains \cite{7914672, 9815098}.

\par
In addition to the pursuit of accurate channel estimation, computational efficient channel estimation methods for massive MIMO systems are also desired to reduce the latency and cope with potential channel variations. Some representative model-driven CP-based estimation methods for a MIMO channel or its latent parameters are listed in Table \ref{tab:t1}. Following array signal processing, the spatial MIMO channel can be constructed by collecting the steering vectors of the different paths/users, thereby leading to a Vandermonde matrix \cite{van2002optimum}. Hence, for tensor MIMO channels, the CP factors could also be restricted by such a Vandermonde structure, which results in Vandermonde-constrained CP (VCP) decomposition \cite{6573422}. Furthermore, by exploiting the Vandermonde structure of the factors, subspace-type algorithms based on the singular value decomposition (SVD) and estimation of signal parameters via rotational invariance techniques (ESPRIT) are developed in \cite{32276}. Various subspace-type estimation methods to calculate Vandermonde-constrained CP decomposition in different MIMO systems are proposed in \cite{9792305, 9835123, 9049103, 9540372,zhang2024integrated}.
There are different methods for calculating the CP decomposition, such as gradient descent, quasi-Newton and nonlinear least squares \cite{sidiropoulos2017tensor, sorber2013optimization}. Nevertheless, the alternating least squares (ALS) is described as the ``workhorse'' for calculating the CP decomposition \cite{kolda2009tensor}, and is widely used in massive MIMO channel estimation \cite{8922606, 8481590, 7556337, 9606606, 8693958, 9361077, 7914672, 9815098, 8895800,10416244}. The ALS is conceptually very simple, and only updates one factor at the time using least squares minimization. Thus, it can always reduce the objective function value monotonically. Many early works point out that the convergence speed and converged stationary points of the ALS depend on the initialization of the algorithm, and good starting points can help to speed up the ALS and find the minima \cite{smilde2005multi, sanchez1990tensorial, bro1997parafac}. While some competing methods may produce superior performance, the ALS can chase them once given better initializations \cite{sorber2013optimization}. However, the existing CPALS methods for channel estimation mostly use random initialization and SVD based initialization \cite{kolda2009tensor}, resulting in thousands of iterations to achieve an expected estimation accuracy, which reduces their practicality. The reason for this problem is that existing initialization strategies are decoupled from the ALS iterations, and are not designed for converging quickly and well, which inspires us to formulate the initialization and iteration into an end-to-end framework.

\begin{table}[t!]
\renewcommand\arraystretch{1.05}
\caption{A list of model-driven CP-based estimation methods for a MIMO channel or its latent parameters $^\ddag$. }
\label{tab:t1}
\begin{center}
\begin{tabular}{cc}
\toprule
MIMO Systems & Algorithms \\
\midrule
  MIMO radar \cite{8922606} & ALS \\
  MU MIMO\cite{7556337} & ALS \\
  DP MIMO \cite{8481590} & ALS \\
  CF MIMO \cite{9792305} & subspace-type \\
  TV MIMO \cite{9606606, 8693958} & ALS \\
  TV MU MIMO \cite{9835123} & subspace-type \\
  MIMO-OFDM \cite{7914672} & ALS \\
  DWB MIMO-OFDM \cite{9049103} & subspace-type \\
  TV MIMO-OFDM \cite{9815098, 8895800} &  ALS \\
  IRS MIMO \cite{9361077,10137372, 9540372} & subspace-type/ALS \\  
  ISAC MIMO \cite{zhang2024integrated,10416244, 10643882} & subspace-type/ALS \\
  MA-enabled MIMO \cite{zhang2024channel} & ALS \\
\bottomrule
\end{tabular}
\end{center}
\footnotesize{$^\ddag$ MU: multi-user. DP: dual-polarized. CF: cell-free. TV: time-varying. DWB: dual-wide-band. IRS: intelligent-reflecting-surface-assisted. ISAC: integrated sensing and communication.}\\
\end{table}

\par
Recently, deep neural networks (DNNs) show a powerful capability to capture the wireless channel characteristics from tons of data, and they have been widely applied for MIMO channel estimation \cite{ye2017power, 8640815, 8400482}. By treating the channel estimation as a denoising task, the spatial-frequency convolutional neural network (SFCNN) has been proposed for multidimensional channel estimation, which uses a convolutional neural network (CNN) to exploit the multi-dimensional spatial and frequency structure of a MIMO-OFDM channel \cite{dong2019deep}. Unfortunately, DNNs are commonly utilized as a black box and data-driven deep learning does not yet offer the interpretability and reliability of model-based methods \cite{9363511}.
As an alternate that benefits from the advantages of both model-driven and data-driven paradigms, model-based deep learning methods \cite{10056957} have attracted the attention of the MIMO communication research, which generally incorporate an internal or external DNN into an iterative algorithm. Compared with traditional algorithms implemented for an individual sample, they benefit from learning the domain knowledge and have shown a performance improvement while keeping relative interpretability \cite{7934066, he2018deep, 9298921}.
However, these methods are designed for matrix signals and are barely applied to exploit the multidimensional tensor signal structure. Focusing on the tensor network decomposition, a core tensor network (CTN) is proposed in \cite{9413637} that integrates the gradient-descent-based tensor decomposition and transfer learning to learn a mapping of correlated tensors, which improves the initial condition for gradient-descent.

\par

In this paper, we propose a deep-learning-aided CPALS (DL-CPALS) for CP decomposition and massive MIMO tensor channel estimation, which integrates the unrolled model-driven low-rank approximation algorithm with the deep-learning process to enhance the speed and accuracy. The contributions and advantages of this work are summarised as follows:
\begin{itemize}
\item A novel DL-aided tensor low-rank decomposition framework is introduced, which could be easily applied to other tensor decomposition models. As long as the algorithm is differentiable to ensure a gradient-based back-propagation, the CPALS of the framework can be replaced by alternative tensor decomposition algorithms.
\item An easy-to-implement DNN is developed for the channel estimation application. Specifically, the DL-CPALS augments the workhorse algorithm CPALS by using a simple fully connected neural network to encode an input tensor into the initializations for the unrolled ALS using a small number of iterations. By integrating the DNN-based initialization and the CPALS algorithm into an end-to-end formulation, the DNN is trained to learn good initializations that are favorable for achieving a small low-rank approximation loss efficiently with CPALS.
\item Different from existing DNNs for channel denoising, as a DL-aided model-driven method, the performance of the proposed method is guaranteed as it represents iterations of the CPALS. Moreover, benefiting from the denoising effect of a low-rank approximation of the CPALS itself, the proposed method is trained by noisy data and does not require the corresponding real clean data.
\end{itemize}
Compared with existing initialization strategies, experimental results demonstrate that the proposed method leads to a more accurate CP decomposition and channel estimate by using fewer iterative steps.


\section{Notations}
Tensors, matrices, vectors and scalars are represented by boldface calligraphic uppercase letters $\boldsymbol{\mathcal{X}}$, boldface capital letters $\mathbf{X}$, boldface lower case letters $\mathbf{x}$ and lowercase letters $x$, respectively. The symbol $j$ is used to represent $\sqrt{-1}$. Superscripts $(\cdot)^T, (\cdot)^*, (\cdot)^H$ and $(\cdot)^\dag$ denote the transpose, complex conjugate, Hermitian transpose and pseudoinverse, respectively. Symbols $\circ$, $\odot$, $\diamond$ and $\otimes$ denote the outer product, Hadamard product, Khatri-Rao product and Kronecker product, respectively. The operators $\text{rank}(\cdot)$ and $\text{krank}(\cdot)$ denote the matrix rank and Kruskal rank, respectively. The operator $\text{diag}(\cdot)$ converts a vector into a diagonal matrix. $\mathcal{U}(a,b)$ denotes the uniform distribution in the interval $(a,b)$, and $\mathcal{CN}(m,\sigma^2)$ denotes the complex circularly-symmetric Gaussian distribution with mean value $m$ and variance $\sigma^2$. $\mathbf{1}_{m}\in\mathbb{R}^{m}$ denotes the all-one vector.
Mode-$n$ matricization of a tensor $\boldsymbol{\mathcal{X}}\in\mathbb{C}^{I_1\times ... \times I_n\times ...\times I_N}$ is denoted as $\mathbf{X}_{(n)}\in\mathbb{C}^{I_n\times \prod_{m\neq n}^N I_m}$, where index $n$ enumerates the rows and the rest enumerates the columns.
The mode-$n$ product of the tensor $\boldsymbol{\mathcal{X}}$ and a matrix $\mathbf{A}_n\in\mathbb{R}^{K \times I_n}$ is a tensor $\boldsymbol{\mathcal{X}}\times_n \mathbf{A}_n\in\mathbb{R}^{I_1\times \ldots \times I_{n-1}\times K\times I_{n+1}\ldots \times I_N}$, whose elements are computed by $[\boldsymbol{\mathcal{X}}\times_n \mathbf{A}_n]_{i_1, \ldots, i_{n-1},k,i_{n+1}, \ldots, i_N}=\sum_{i_n = 1}^{I_n}x_{i_1, \ldots,i_n,\ldots, i_N}a_{k,i_n}$.
Denote $\Diamond^N_{l\neq n}\mathbf{A}_l=\mathbf{A}_1\diamond\ldots\mathbf{A}_{n-1}\diamond\mathbf{A}_{n+1}\diamond\ldots\mathbf{A}_{N}$ as the Khatri-Rao product of all but one matrix. Similarly, denote $\bigodot^N_{l\neq n}\mathbf{A}_l=\mathbf{A}_1\odot\ldots\mathbf{A}_{n-1}\odot\mathbf{A}_{n+1}\odot\ldots\mathbf{A}_{N}$ as the Hadamard product of all but one matrix. A matrix $\mathbf{A}\in\mathbb{C}^{I\times R}$ is said to be (exponential) Vandermonde if its elements are $[\mathbf{A}]_{i,r}=e^{-2\pi j(i-1)\omega_{r}}$. Denote $\mathbf{A}=\text{Van}(\mathbf{z})$ as the generation of a Vandermonde matrix $\mathbf{A}$ using $\mathbf{z}=2\pi[\omega_{1},\ldots,\omega_{R}]^T$, where $\mathbf{z}$ is called the generating vector.

\section{Preliminaries}
\par
Here we introduce the CP decomposition and the ALS algorithm applied on tensor channel estimation.
For a tensor $\boldsymbol{\mathcal{X}}\in\mathbb{C}^{I_1\times \ldots \times I_N}$, its CP decomposition is formulated as
\begin{equation}
\begin{split}
\boldsymbol{\mathcal{X}}= \sum_{r=1}^R \alpha_r \mathbf{a}_{1, r}\circ\ldots\circ\mathbf{a}_{N, r}
=\llbracket \boldsymbol{\alpha}; \mathbf{A}_1, \ldots, \mathbf{A}_N\rrbracket
,
\end{split}
\label{cp1}
\end{equation}
where $\alpha_r$ is the weight of the $r$th rank-one component, $\llbracket \cdot \rrbracket$ denotes the simplified CP representation and $\boldsymbol{\alpha}=[\alpha_1,\dots,\alpha_R]^T$. $\mathbf{A}_{n}=[\mathbf{a}_{n, 1},\ldots,\mathbf{a}_{n, R}]$ denotes the factor along the $n$th mode. The CP rank is defined as the smallest value of $R$ satisfying equation (\ref{cp1}). $\boldsymbol{\mathcal{X}}$ is said to be CP low-rank if $R$ is relatively small. Then the high-dimensional $\boldsymbol{\mathcal{X}}$ can be modeled in a low-dimensional parameter space \cite{hong2020generalized}, i.e., $R\sum_{n=1}^N I_n \ll \prod_{n=1}^N I_n$.
The mode-$n$ matricization of a CP tensor is expressed as $\mathbf{X}_{(n)}=\mathbf{A}_n\mathbf{\Lambda}(\mbox{$\Diamond^N_{l\neq n}\mathbf{A}_l$})^T$,
where $\mathbf{\Lambda}=\text{diag}(\boldsymbol{\alpha})$.
The scaling and permutation ambiguities always exist in the CP model. Specifically, if there are some scaling diagonal matrices $\{\boldsymbol{\Delta}_n\}_{n=1}^N$ and a permutation matrix $\boldsymbol{\Upsilon}$ that map the factors $\{\mathbf{A}_n\}_{n=1}^N$ of (\ref{cp1}) as
\begin{equation}
\begin{split}
&\hat{\mathbf{A}}_n = \mathbf{A}_n\boldsymbol{\Delta}_n\boldsymbol{\Upsilon},\ \text{for}\ n=1,\ldots,N,
\\
&\prod_{n=1}^N\boldsymbol{\Delta}_n= \text{diag}(\boldsymbol{\Upsilon}\boldsymbol{\alpha}),
\end{split}
\label{cpambiguity}
\end{equation}
then we have
\begin{equation}
\begin{split}
\boldsymbol{\mathcal{X}}= \llbracket \boldsymbol{\alpha}; \mathbf{A}_1, \ldots, \mathbf{A}_N\rrbracket = \llbracket \mathbf{1}_{R}; \hat{\mathbf{A}}_1, \ldots, \hat{\mathbf{A}}_N\rrbracket.
\end{split}
\end{equation}
Note that the uniqueness of the CP decomposition does not consider such ambiguity. The following theorem \cite{CPuniqueness} gives a sufficient condition for uniqueness.
\begin{theo}[Uniqueness condition\cite{CPuniqueness}]
Let an $N$th-order tensor $\boldsymbol{\mathcal{X}}$ satisfy (\ref{cp1}). Then, the CP decomposition is unique if
\begin{equation}
\begin{split}
\sum_{n=1}^N \text{krank}(\mathbf{A}_n) \geq 2R + N - 1.
\end{split}
\end{equation}
\label{The1}
\end{theo}
\par
When the CP model is applied to the received tensor signals of MIMO communication systems equipped with uniform linear arrays (ULAs) or uniform rectangular arrays (URAs), part of the factors may become Vandermonde matrices, which leads to a VCP model \cite{6573422}, i.e., $
\boldsymbol{\mathcal{X}}= \llbracket \boldsymbol{\alpha}; \mathbf{A}_1, \ldots, \mathbf{A}_{N_1}, \mathbf{A}_{N_1 + 1}, \ldots,  \mathbf{A}_{N}\rrbracket$,
where the factors $\{\mathbf{A}_{n}\}_{n = N_1 + 1}^{N}$ are Vandermonde matrices whose elements can be formulated as $[\mathbf{A}_{n}]_{i_n,r}=e^{-2\pi j(i_n-1)\omega_{n, r}}$, and where the factors $\{\mathbf{A}_{n}\}_{n = 1}^{N_1}$ are generalized matrices with no Vandermonde structure. The generating vectors of the Vandermonde factors can carry information about the AoAs, AoDs, delays and so on, which depends on the specific system as summarized in Table \ref{tab:t1}.
\par
Without accounting for the Vandermonde structure, the ALS algorithm can be used for calculating the VCP decomposition by solving the optimization problem as
\begin{equation}
\begin{split}
\min_{\{\hat{\mathbf{A}}_n\}^N_{n=1}}
\left\|\boldsymbol{\mathcal{Y}} - \llbracket \mathbf{1}_{R}; \hat{\mathbf{A}}_1, \ldots, \hat{\mathbf{A}}_N\rrbracket\right\|^2_F.
\end{split}
\label{ycpd}
\end{equation}
The ALS updates factor $\hat{\mathbf{A}}_n$ alternatingly by fixing the other factors. Thus, the optimization problem (\ref{ycpd}) in the mode-$n$ matricization form becomes
\begin{equation}
\begin{split}
\min_{\hat{\mathbf{A}}_n}
\left\|\mathbf{Y}_{(n)} -  \hat{\mathbf{A}}_n \left(\mbox{$\Diamond^N_{l\neq n}\hat{\mathbf{A}}_l$}\right)^T \right\|^2_F,
\end{split}
\label{ycpd-step}
\end{equation}
and its LS solution is given as
\begin{equation}
\begin{split}
\hat{\mathbf{A}}_n&=\mathbf{Y}_{(n)}\left(\mbox{$\Diamond^N_{l\neq n}\hat{\mathbf{A}}_l$}\right)^{T, \dag}
\\
&=\mathbf{Y}_{(n)}\left(\mbox{${\Diamond}^N_{l\neq n}\hat{\mathbf{A}}_{l}^*$}\right)\left(\bigodot^N_{l\neq n} \hat{\mathbf{A}}^{T}_{l}\hat{\mathbf{A}}^{*}_{l}\right)^{\dag}.
\end{split}
\label{als-step}
\end{equation}
We summarize the CPALS in Algorithm \ref{alg:1}. Note that the tolerance on the difference of the factors or reconstructed tensor between two adjacent iterations can also be used as the terminating condition. Nevertheless, we use the number of iterations $K$ as the terminating condition to ease description. To simplify notations, we denote $\Phi$ as the operator to calculate CPALS in Algorithm \ref{alg:1}, which is formulated as
\begin{equation}
\begin{split}
\{\hat{\mathbf{A}}^K_{n}\}_{n=1}^N = \Phi_{R, K}(\boldsymbol{\mathcal{Y}}, \{\hat{\mathbf{A}}^0_{n}\}_{n=2}^N),
\end{split}
\label{als}
\end{equation}
where $\boldsymbol{\mathcal{Y}}$ is the input tensor that needs to be decomposed and $\hat{\mathbf{A}}^0_{n}$ is the initial factor for $n=2,\ldots,N$. The approximated CP rank $R$ and the number of iterations $K$ are treated as the predefined hyper-parameters of $\Phi$.

\begin{algorithm}[t!]
\renewcommand{\algorithmicrequire}{\textbf{Input:}}
\renewcommand\algorithmicensure {\textbf{Output:} }
\caption{CPALS $\Phi_{R, K}(\boldsymbol{\mathcal{Y}}, \{\hat{\mathbf{A}}^0_{n}\}_{n=2}^N)$}
\label{alg:1}
\begin{algorithmic}[1]
\REQUIRE tensor $\boldsymbol{\mathcal{Y}}$, initializations $\{\hat{\mathbf{A}}^0_{n}\}_{n=2}^N$, CP rank $R$ and the number of iterations $K$
\ENSURE  updated factors $\{\hat{\mathbf{A}}^K_{n}\}_{n=1}^N$
\STATE Update $\hat{\mathbf{A}}^0_1\leftarrow\mathbf{Y}_{(1)}\left(\mbox{${\Diamond}^N_{n=2}\hat{\mathbf{A}}_{n}^{0,*}$}\right)
\left(\bigodot^N_{n=2} \hat{\mathbf{A}}^{0,T}_{n}\hat{\mathbf{A}}^{0,*}_{n}\right)^{\dag}$.
\FOR {$k=1:K$}
\FOR {$n=1:N$}
\STATE Update $\hat{\mathbf{A}}^k_{n}$ by calculating (\ref{als-step}).
\ENDFOR
\ENDFOR
\end{algorithmic}
\end{algorithm}

\par
After the ALS terminates, we can extract the $r$th elements of the generating vectors $\{\hat{\mathbf{z}}_n\}_{n = N_1 + 1}^{N}$ of the Vandermonde factors by solving the maximum correlation problem
\begin{equation}
\begin{split}
\max_{\hat{z}_{n, r}}
\frac{\left|\hat{\mathbf{a}}_{n,r}^H\text{Van}(\hat{z}_{n, r})\right|}
{\left\|\hat{\mathbf{a}}_{n,r}\right\|_2},
\end{split}
\label{est_z}
\end{equation}
where $\hat{\mathbf{a}}_{n,r}$ is the $r$th column of $\hat{\mathbf{A}}_n$. The above problem can be solved by a simple one-dimensional search method.

\section{Deep-learning-aided CP decomposition}
In this section, we integrate the CP decomposition into the data-driven learning process and propose the deep-learning-aided CPALS, where we use a DNN to generate the initializations that are favorable for CP decomposition. Hence, the proposed method enhances the algorithm's accuracy and efficiency.

\subsection{Motivations}
\label{sec:motivation}
\begin{figure}[t!]
\begin{center}
\includegraphics[width=1\linewidth]{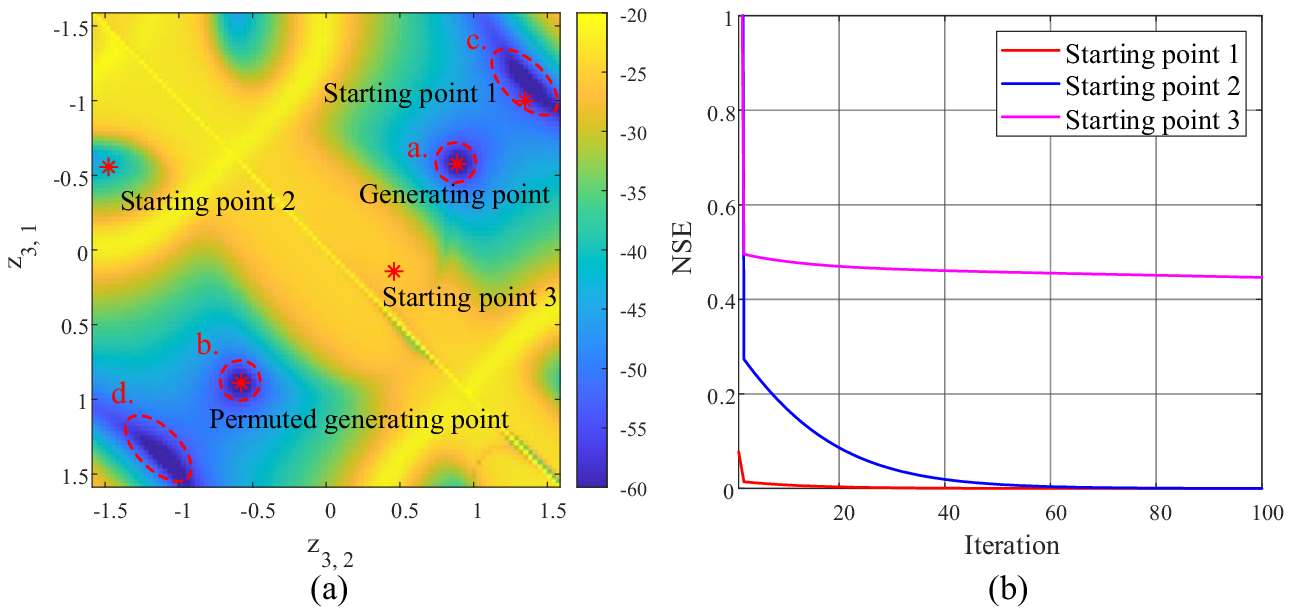}
\end{center}
\caption{Iterative behaviour of ALS using different starting points of $\hat{\mathbf{z}}^0_3=[z_{3,1}, z_{3,2}]^T$ for computing a random tensor $\boldsymbol{\mathcal{Y}}$. (a) The reconstruction NSE of $\frac{\left\|\hat{\boldsymbol{\mathcal{Y}}}-\boldsymbol{\mathcal{Y}}\right\|^2_F}{\left\|\boldsymbol{\mathcal{Y}}\right\|^2_F}$ in dB with $K=50$ using different starting points, where $\hat{\boldsymbol{\mathcal{Y}}}$ is the reconstructed tensor. (b) The reconstruction NSE as a function of iteration number.}
\label{fig:Fig1}
\end{figure}
Researchers have long found in experiments that unfavorable starting points can bring CPALS into swamps \cite{mitchell1994slowly, smilde2005multi}, where CPALS has not converged to a minimum and using more iterations would not further reduce the objective value. Favorable initializations not only speed up CPALS in finding a mimimum, but also can ensure the global or local convergence theoretically \cite{uschmajew2012local, wang2014global}. As the initialization plays an important role for CPALS, two questions naturally arise. \textit{What are favorable initializations for CPALS}? And \textit{how to generate favorable initializations}? Next, we address these two questions as a way to motivate our proposed approach.

\par
\textit{What are favorable initializations for CPALS}? To answer this question, we first illustrate the iterative behaviour of CPALS using different initializations. Consider a synthetic VCP tensor $\boldsymbol{\mathcal{Y}}\in\mathbb{C}^{4\times 4 \times 4}$ with $R=2$, whose factors are all Vandermonde matrices. Elements of the generating vectors are drawn from $\mathcal{U}(-\frac{\pi}{2},\frac{\pi}{2})$ randomly, which are $\mathbf{z}_1=[0.36, 0.18]^T$, $\mathbf{z}_2=[1.10, -0.70]^T$ and $\mathbf{z}_3=[-0.58, 0.89]^T$. Note that this tensor satisfies the uniqueness condition of Theorem \ref{The1}. Here we only consider to decompose $\boldsymbol{\mathcal{Y}}$ with CPALS without calculating (\ref{est_z}) for extracting parameters. We need to provide initializations of $\hat{\mathbf{z}}^0_2$ and $\hat{\mathbf{z}}^0_3$ to construct the initialization factors $\{\hat{\mathbf{A}}^0_{n}\}_{n=2}^3$, as shown in Algorithm 1. We use the randomly generated initialization $\hat{\mathbf{z}}^0_2=[-0.46, -0.85]^T$, and evaluate the performance of CPALS with different starting points of $\hat{\mathbf{z}}^0_3=[z_{3,1}, z_{3,2}]^T$. In Figure \ref{fig:Fig1} (a) and \ref{fig:Fig1} (b), we show the normalized square error (NSE) calculated by $\frac{\left\|\hat{\boldsymbol{\mathcal{Y}}}-\boldsymbol{\mathcal{Y}}\right\|^2_F}{\left\|\boldsymbol{\mathcal{Y}}\right\|^2_F}$, where $\hat{\boldsymbol{\mathcal{Y}}}=\llbracket \mathbf{1}_{R}; \hat{\mathbf{A}}^K_1, \hat{\mathbf{A}}^K_2, \hat{\mathbf{A}}^K_3\rrbracket$ is the reconstructed tensor by using CPALS. As shown in Figure \ref{fig:Fig1} (a), four regions, which are surrounded by dashed circles, give favorable initializations that allow CPALS to quickly reduce the estimation error to 0. The first two regions, i.e., region a and region b, are near the true generating point $\mathbf{z}_3=[0.89, -0.58]^T$ and its permuted point $[-0.58, 0.89]^T$ due to the permutation ambiguity. Interestingly, there are two more such regions, i.e., region c near the starting point 1 and the permuted region d.
In Figure \ref{fig:Fig1} (b), we plot the NSE as a function of the iteration number for CPALS by using the selected starting points.
We can observe that the objective value using starting point 3 reduces slowly like in a swamp, but it reduces quickly using starting point 1. Hence, we could describe that favorable initializations would make the ALS reduce to a low objective value or decomposed error using just a few iterations. Note that when $\boldsymbol{\mathcal{Y}}$ is disturbed by the noise $\boldsymbol{\mathcal{N}}$, a low objective value of the approximation problem in (\ref{ycpd}), i.e., $\left\|\hat{\boldsymbol{\mathcal{Y}}}-\boldsymbol{\mathcal{Y}}- \boldsymbol{\mathcal{N}}\right\|^2_F$, may not ensure a low NSE. Nevertheless, using a small objective value as the guidance is feasible if the uniqueness condition is satisfied and the noise is small.

\par
\textit{How to generate favorable initializations}? We denote $\Psi(\boldsymbol{\mathcal{Y}})$ as a functional generator that produces initializations for CPALS to decompose $\boldsymbol{\mathcal{Y}}$. In addition to the random initialization, the most commonly used strategy is based on the SVD, i.e.,
\begin{equation}
\begin{split}
\Psi(\boldsymbol{\mathcal{Y}}): \ &\hat{\mathbf{A}}^0_{n} = R\ \text{leading left singular vectors of}\ \mathbf{Y}_{(n)},\\
&\text{for } n = 2,\ldots, N.
\end{split}
\label{svdini}
\end{equation}
However, this initialization strategy fails to consider the updating steps of CPALS in the sequel and does not ensure to generate favorable initializations. Although one can also use other non-iterative decomposition methods like the generalized eigenvalue decomposition (GEVD) to pre-estimate a starting point close to the ground truth as initialization \cite{vervliet2016tensorlab}, they have higher computational complexity than using the SVD.
We need a $\Psi$ providing a guidance that prompts an accurate and fast CP decomposition by producing favorable initializations for the input tensor. Thus, $\Psi$ needs outer domain knowledge learned from a large number of tensor decomposition processes, which inspires us to employ a DNN.

\par

\subsection{The proposed deep-learning-aided CPALS (DL-CPALS)}

We consider to develop a DNN based generator $\Psi_{\boldsymbol{\theta}}$ with parameters $\boldsymbol{\theta}$, which learns to produce favorable initial parameters to decompose new tensors fast and well. Note that the CPALS for a tensor is naturally treated as a model-driven learning task, which updates its model parameters, i.e., the CP factors, starting from the initializations. In this way, our goal is to pursue favorable initializations using a learned initialization generator $\Psi_{\boldsymbol{\theta}}$. Given a batch of CPALS tasks for decomposing tensors $\boldsymbol{\mathcal{Y}}_1,\ldots,\boldsymbol{\mathcal{Y}}_{L_{\text{tr}}}$, we can integrate them into the data-driven learning process of $\Psi_{\boldsymbol{\theta}}$. If the unrolled CPALS is differentiable for updating the CP factors, gradients can be backpropagated to train $\Psi_{\boldsymbol{\theta}}$ for generating favorable initializations. Specifically, the loss function of the learning process is formulated as
\begin{equation}
\begin{split}
\min_{\boldsymbol{\theta}}&\ \frac{1}{L_{\text{tr}}}\sum_{l=1}^{L_{\text{tr}}}\left \|\boldsymbol{\mathcal{Y}}_l-\llbracket \mathbf{1}_{R}; \hat{\mathbf{A}}^K_{l, 1}, \ldots, \hat{\mathbf{A}}^K_{l, N}\rrbracket\right\|_F^2,
\\
\text{s.t.}\ &\ \{\hat{\mathbf{A}}^0_{l, n}\}_{n=2}^{N}=\Psi_{\boldsymbol{\theta}}(\boldsymbol{\mathcal{Y}}_l),\ \forall l,
\\
&\ \{\hat{\mathbf{A}}^K_{l, n}\}_{n=1}^N = \Phi_{R, K}(\boldsymbol{\mathcal{Y}}_l, \{\hat{\mathbf{A}}^0_{l, n}\}_{n=2}^N),\ \forall l.
\end{split}
\label{DLALS}
\end{equation}
Note that the loss matches the objective function of the CP low-rank approximation problem in (\ref{ycpd}). When we set a small $K$, the proposed method aims to optimize a DNN for generating the initializations such that a small number of ALS iterations on a CP decomposition task will produce a maximally effective behavior to reduce the approximation error.
Thus, the learned DNN would be a good initialization generator that can let the ALS converge to a low objective error quickly for decomposing an input tensor. Moreover, as a model-based optimization algorithm, CPALS is able to find a clean low-rank tensor from a noisy high-rank tensor by solving the low-rank approximation problem, which leads to a denoising effect. Hence, optimizing the loss function (\ref{DLALS}) and training the DNN $\Psi_{\boldsymbol{\theta}}$ does not need the paired clean data. The training data $\boldsymbol{\mathcal{Y}}_1,\ldots,\boldsymbol{\mathcal{Y}}_{L_{\text{tr}}}$ can be noisy, which improves the practicability of the proposed method, as clean data are not available in many cases, e.g., wireless channel responses.

\begin{algorithm}[t!]
\renewcommand{\algorithmicrequire}{\textbf{Input:}}
\renewcommand\algorithmicensure {\textbf{Output:}}
\caption{Training of DL-CPALS}
\label{alg:dlcpals}
\begin{algorithmic}[1]
\REQUIRE the tensor dataset $\{\boldsymbol{\mathcal{Y}}_{l}\}_{l=1}^{L_{\text{tr}}}$ obeying $p(\boldsymbol{\mathcal{Y}})$, the CP rank $R$, the number of iterations $K$, the number of epochs $N_{\text{epoch}}$ and the tensor mini-batch size $L$
\ENSURE  DNN $\Psi_{\boldsymbol{\theta}}$
\FOR {$i=1:N_{\text{epoch}}$}
\FOR {$j=1:L_{\text{tr}}/L$}
\STATE Draw $L$ tensors from dataset without replacement.
\FOR {$l=1:L$}
\STATE Input $\boldsymbol{\mathcal{Y}}_l$ into $\Psi_{\boldsymbol{\theta}}$ and obtain the initialization factors $\{\hat{\mathbf{A}}^0_{l, n}\}_{n=2}^{N}=\Psi_{\boldsymbol{\theta}}(\boldsymbol{\mathcal{Y}}_l)$.
\STATE Carry out CPALS and obtain updated factors $\{\hat{\mathbf{A}}^K_{l, n}\}_{n=1}^N = \Phi_{R, K}(\boldsymbol{\mathcal{Y}}_l, \{\hat{\mathbf{A}}^0_{l, n}\}_{n=2}^N)$.
\STATE Calculate the loss of low-rank CP approximation, i.e., $\mathcal{L}_{l} = \left \|\boldsymbol{\mathcal{Y}}_l-\llbracket \mathbf{1}_{R}; \hat{\mathbf{A}}^K_{l, 1}, \ldots, \hat{\mathbf{A}}^K_{l, N}\rrbracket\right\|_F^2$
\ENDFOR
\STATE Calculate the mini-batch-wise loss $\frac{1}{L}\sum_{l=1}^L\mathcal{L}_{l}$, and update $\boldsymbol{\theta}$ using optimizer.
\ENDFOR
\ENDFOR
\end{algorithmic}
\end{algorithm}

We provide the training steps of DL-CPALS in Algorithm \ref{alg:dlcpals}, where we use a batch-wise training strategy. A prerequisite to optimize the loss function (\ref{DLALS}) is that the CPALS $\Phi_{R, K}$ is differentiable in the updated factors for all tensors, which ensures the gradient back-propagation in line 6 of Algorithm \ref{alg:dlcpals}.
As shown in (\ref{als-step}), the iterative updating steps of $\Phi_{R, K}$ employ a multiplication, Khatri-Rao product, Hadamard product and pseudoinverse of matrices, which are all differentiable \cite{hjorungnes2007complex}. The open source deep learning framework PyTorch \cite{paszke2019pytorch} is able to compute gradients of these matrix operations automatically\footnote{https://pytorch.org/docs/stable/generated/torch.linalg.pinv.html}.
If $\bigodot^N_{l\neq n} \hat{\mathbf{A}}^{T}_{l}\hat{\mathbf{A}}^{*}_{l}$ has full rank, its pseudoinverse is equal to the inverse. Using the inverse operation can simplify the gradient calculation and make the gradient numerically stable. As we find that $\bigodot^N_{l\neq n} \hat{\mathbf{A}}^{T}_{l}\hat{\mathbf{A}}^{*}_{l}$ is usually full rank in experiments, we suggest using the inverse operation. Because the derivatives of $\Phi_{R, K}$ are complicated, here we only formulate the updating of $\hat{\mathbf{A}}^0_{1}$ with $N=3$ for instance, which is processed as
\begin{equation}
\begin{split}
\hat{\mathbf{A}}^0_{1} &= \Phi_{R, 0}(\boldsymbol{\mathcal{Y}}, \{\hat{\mathbf{A}}^0_{2},\hat{\mathbf{A}}^0_{3}\})
\\
&=\mathbf{Y}_{(1)}\left(\hat{\mathbf{A}}_{2}^{0,*}{\diamond}\hat{\mathbf{A}}_{3}^{0,*}\right)\left( \hat{\mathbf{A}}^{0,T}_{2}\hat{\mathbf{A}}^{0,*}_{2} \odot \hat{\mathbf{A}}^{0,T}_{3}\hat{\mathbf{A}}^{0,*}_{3}\right)^{-1}.
\end{split}
\end{equation}
Due to $\{\hat{\mathbf{A}}^0_{2},\hat{\mathbf{A}}^0_{3}\}=\Psi_{\boldsymbol{\theta}}(\boldsymbol{\mathcal{Y}})$, both $\hat{\mathbf{A}}^0_{2}\in\mathbb{C}^{I_2\times R}$ and $\hat{\mathbf{A}}^0_{3}\in\mathbb{C}^{I_3\times R}$ are related to the DNN parameters $\boldsymbol{\theta}$. For ease of formulation, we only consider $\hat{\mathbf{A}}_{2}^{0}$ as a variable matrix and fix $\hat{\mathbf{A}}_{3}^{0}$ as a constant. 
Let
\begin{equation}
\begin{split}
\mathbf{g}(\text{vec}(\hat{\mathbf{A}}_{2}^{0})) &= \text{vec}\left(\left( \hat{\mathbf{A}}^{0,T}_{2}\hat{\mathbf{A}}^{0,*}_{2} \odot \hat{\mathbf{A}}^{0,T}_{3}\hat{\mathbf{A}}^{0,*}_{3}\right)^{-1}\right)\in\mathbb{C}^{R^2}
\end{split}
\label{Dfg}
\end{equation}
represent a matrix function of $\hat{\mathbf{A}}_{2}^{0}$, where $\text{vec}(\cdot)$ is the vectorization operator that stacks the column vectors of the argument matrix into
a long column vector. The matrix function $\mathbf{g}$ maps an $I_2 R$-dimensional variable to the $R^2$-dimensional space. We define the derivative as $\mathcal{D}_{\text{vec}(\hat{\mathbf{A}}^0_{2})} \mathbf{g}\in\mathbb{C}^{ R^2 \times I_2 R}$.
First, we have
\begin{equation}
\begin{split}
\text{vec}\left(\hat{\mathbf{A}}^{0,T}_{2}\hat{\mathbf{A}}^{0,*}_{2}\right)=\left(\hat{\mathbf{A}}^{0,H}_{2}\otimes \mathbf{I}_R\right)\text{vec}\left(\hat{\mathbf{A}}^{0,T}_{2}\right).
\end{split}
\end{equation}
Thus, we can obtain
\begin{equation}
\begin{split}
\mathcal{D}_{\text{vec}(\hat{\mathbf{A}}^0_{2})} \text{vec}\left(\hat{\mathbf{A}}^{0,T}_{2}\hat{\mathbf{A}}^{0,*}_{2}\right)=\left(\hat{\mathbf{A}}^{0,H}_{2}\otimes \mathbf{I}_R\right)\mathbf{K}_{I_2, R},
\end{split}
\label{DA}
\end{equation}
where $\mathbf{K}_{k, l}\in\mathbb{R}^{kl\times kl}$ is called the commutation matrix satisfying $\mathbf{K}_{k, l}\text{vec}(\mathbf{Z})=\text{vec}(\mathbf{Z}^T)$ for the matrix $\mathbf{Z}\in\mathbb{C}^{k\times l}$.
According to [48], two important expressions are
\begin{equation}
\begin{split}
&\mathcal{D}_{\text{vec}(\mathbf{X})} \text{vec}\left(\mathbf{X}^{-1}\right) = -\mathbf{X}^{-T}\otimes \mathbf{X}^{-1},
\\
&\mathcal{D}_{\text{vec}(\mathbf{X})} \text{vec}\left(\mathbf{X}\odot \mathbf{Y}\right) = \text{diag}(\text{vec}(\mathbf{Y})).
\end{split}
\label{DX}
\end{equation}
By using (\ref{Dfg}), (\ref{DA}) and (\ref{DX}), we can obtain
\begin{equation}
\begin{split}
&\mathcal{D}_{\text{vec}(\hat{\mathbf{A}}^0_{2})}\mathbf{g}
\\
=&-\left(\mathbf{g}^{T}\otimes \mathbf{g}\right)\text{diag}(\text{vec}(\hat{\mathbf{A}}^{0,T}_{3}\hat{\mathbf{A}}^{0,*}_{3}))\left(\hat{\mathbf{A}}^{0,H}_{2}\otimes \mathbf{I}_R\right)\mathbf{K}_{I_2, R}.
\end{split}
\end{equation}
Then, we can derive
\begin{equation}
\begin{split}
\mathcal{D}_{\text{vec}(\hat{\mathbf{A}}^0_{2})} \text{vec}(\Phi_{R, 0}) =&  \left(\mathbf{I}_R \otimes \mathbf{Y}_{(1)}\left(\hat{\mathbf{A}}_{2}^{0,*}{\diamond}\hat{\mathbf{A}}_{3}^{0,*}\right)\right)\mathcal{D}_{\text{vec}(\hat{\mathbf{A}}^0_{2})}\mathbf{g}.
\end{split}
\end{equation}
In this way, the differentiable iterative updating steps of CPALS eventually leads to the learnable $\Psi_{\boldsymbol{\theta}}$.

\begin{figure*}[th!]
\begin{center}
\includegraphics[width=0.8\linewidth]{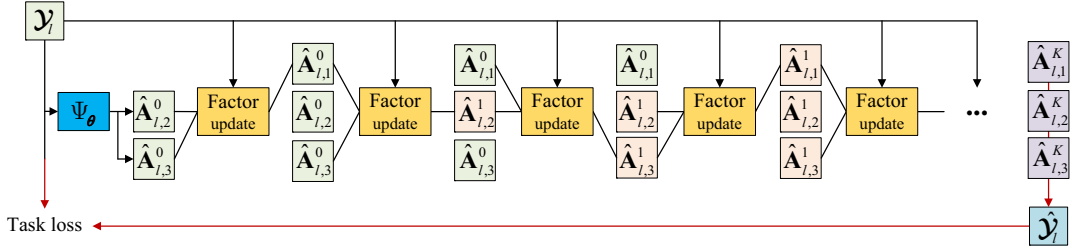}
\end{center}
   \caption{An illustration of the model forward propagation on a single tensor.}
\label{fig:task}
\end{figure*}

Moreover, we illustrate the forward propagation of a $3$rd order tensor in Figure \ref{fig:task}, which corresponds to line 5 and line 6 of Algorithm \ref{alg:dlcpals}. It can be seen that CPALS is unrolled by multiple factor updating steps using (\ref{als-step}). As long as the factor updating steps are differentiable, the DNN parameters $\boldsymbol{\theta}$ can be optimized for learning by gradient back-propagation according to the updating loss in line 9 of Algorithm \ref{alg:dlcpals}.
After the learning terminates, one can obtain the favorable initializations by feeding the test tensor into $\Psi_{\boldsymbol{\theta}}$. Then, performing CPALS can produce an accurate decomposition solution with high efficiency.

\subsection{DNN designs}
\label{sec:DNN}

In this subsection, we describe the details about the network design of the generator $\Psi_{\boldsymbol{\theta}}$. Due to the CP low-rankness assumption that $R$ is small in (\ref{DLALS}), the initializations $\{\hat{\mathbf{A}}^0_{l, n}\}_{n=2}^N$ live in a low-dimensional parameter space in comparison with the reconstructed tensor.
The CPALS $\Phi_{R, K}$ uses a low-dimensional initialization to produce a low-rank approximation of the high-dimensional target tensor, which is like an explicit decoder. Thus, it is natural to design the generator $\Psi_{\boldsymbol{\theta}}$ as an encoder. In addition, $\Psi_{\boldsymbol{\theta}}$ should have a low computational complexity to compete with the SVD-based initialization. We use a simple fully connected network as generator, which outputs initial factors $\{\hat{\mathbf{A}}^0_{l, n}\}_{n=2}^{N}$. Note that the generator $\Psi_{\boldsymbol{\theta}}$ directly outputs the factor matrices rather than generating vectors by utilizing the Vandermonde structure, which makes the proposed method suitable for a generalized CP decomposition.

\par
Given an input tensor $\boldsymbol{\mathcal{Y}}_l\in\mathbb{C}^{I_1\times \ldots \times I_N}$, we reshape it into a vector, and concatenate its real part and imaginary part into a single vector, which has dimension $2\prod_{n=1}^N I_n$. Assume that each hidden layer has $Q$ neurons. Then, the real weight matrices employed on the input layer and the hidden layer are of sizes $2\prod_{n=1}^N I_n \times Q$ and $Q \times Q$, respectively. For the output layer, the weight matrix has size of $Q \times 2R \sum_{n=2}^NI_n$. 
In specific for the MIMO channel estimation application of Section \ref{sec:mimo}, the input dimensions $2 I_1 I_2 M$ and output dimension $2R(I_{1}+M)$ are determined by the numbers of BS antennas $I_2$, MS antennas $I_1$, frequency subcarriers $M$ and paths $R$ of MIMO systems and channels.
We use a Relu activation for the hidden layers. As for the activation of the output layer, we suggest to use the hyperbolic tangent function (Tanh), which constrains the output values in the interval $-1$ to $+1$ so that the energies of the initial factors are not very large and result in imbalance.

\par
If the $\Psi_{\boldsymbol{\theta}}$ is stacked by $D$ hidden layers, considering also the biases, its total number of parameters is $2Q\prod_{n=1}^N I_n+(D-1)Q^2+2QR\sum_{n=2}^NI_n+DQ+2R\sum_{n=2}^NI_n$. If the input is a real tensor, the total number of parameters is $Q\prod_{n=1}^N I_n+(D-1)Q^2+QR\sum_{n=2}^NI_n+DQ+R\sum_{n =2}^NI_n$. In addition, in order to deal with the latent overfitting problem under limited training samples, a dropout layer is added after the activation function of each hidden layer to improve the generalization ability of the DNN model.


\section{Application to massive MIMO-OFDM Tensor Channel Estimation}
\label{sec:mimo}
Although the CPALS can be applied on a variety of massive MIMO systems that are not limited to those listed in Table \ref{tab:t1}, we consider a representative application to MIMO-OFDM tensor channel estimation.
\par
We first consider a downlink MIMO-OFDM system where both the mobile station (MS) and the base station (BS) are equipped with ULAs, which are of size $I_{1}$ and $I_{2}$, respectively. Suppose also that there are in total $M_0$ subcarriers with the sampling rate of $f_{\text{s}}$. 
As described in \cite{7914672, 8306126, dong2019deep}, consider the mmWave channel is formulated based on the geometry-based channel model that has $R$ resolvable paths without clustering.
Denote $\vartheta_r$ and $\theta_r$ as the AoA and AoD of the $r$th path, respectively. To simplify notations, we define $\Theta_{1, r} = 2\pi d\sin\vartheta_r/\lambda$ and $\Theta_{2, r} = 2\pi d\sin\theta_r/\lambda$ as the spatial AoA and AoD.
The MIMO-OFDM channel in the spatial-frequency domain can then be written in CP form as
\begin{equation}
\begin{split}
\boldsymbol{\mathcal{H}}_0
=&\sum_{r=1}^{R} \beta_r e^{-j2\pi \frac{\tau_r f_{\text{s}}}{M_0}} \text{Van}(\Theta_{1, r})\circ\text{Van}(\Theta_{2, r})\circ\text{Van}( \frac{2\pi\tau_r f_{\text{s}}}{M_0})
\\
=&\llbracket \boldsymbol{\beta}; \mathbf{A}_{1}, \mathbf{A}_{2}, \bar{\mathbf{A}}_{3}\rrbracket\in\mathbb{C}^{I_{1} \times I_{2}\times M_0},
\end{split}
\end{equation}
where $\beta_r$ and $\tau_r$ are the equivalent path gain and delay of the $r$th path, respectively. We define $\boldsymbol{\beta}=[\beta_1 e^{-j2\pi \frac{\tau_1 f_{\text{s}}}{M_0}},\ldots,\beta_R e^{-j2\pi \frac{\tau_R f_{\text{s}}}{M_0}}]^T$. We further have $\mathbf{A}_{1}=[\text{Van}(\Theta_{1, 1}),\ldots,\text{Van}_{1}(\Theta_{1, R})]\in\mathbb{C}^{I_{1} \times R}$, $\mathbf{A}_{2}=[\text{Van}_{2}(\Theta_{2, 1}),\ldots,\text{Van}_{2}(\Theta_{2, R})]\in\mathbb{C}^{I_{2}\times R}$,
and $\bar{\mathbf{A}}_{3}=[\text{Van}( \frac{2\pi\tau_1 f_{\text{s}}}{M_0}),\ldots,\text{Van}( \frac{2\pi\tau_R f_{\text{s}}}{M_0})]\in\mathbb{C}^{M_0\times R}$, which are stacked by the steering vectors corresponding to the receiving ULA, transmitting ULA and subcarriers, respectively.

\begin{figure}[!t]
\centering
\includegraphics[width=0.85\linewidth]{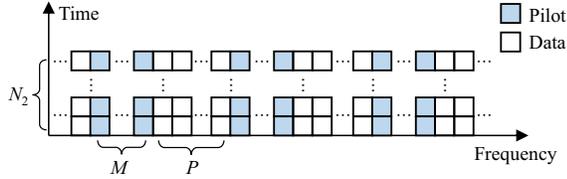}
\caption{An illustration of the pilot pattern.}
\label{fig:pilots}
\vspace{-0.5cm}
\end{figure}

\par
To facilitate the comparison of our method with others in experiments, we follow the pilot transmission protocol described in \cite{dong2019deep}. As the frequency correlation is helpful
to improve the channel estimation accuracy, suppose that a training block has $M$ sequential subcarriers in the frequency domain to transmit $N_2$ training pilots in the time domain, which is illustrated in Figure \ref{fig:pilots}. Two training blocks are separated by $P$ subcarriers dedicated to data transmission. We only consider the estimation of the tensor channel of a training block that corresponds to the pilot positions. The channel is assumed to be block-fading, and it is constant during training. Based on the pilot channels, interpolation can be used to get the data channels. The training tensor channel of the $t$th block is given by
\begin{equation}
\begin{split}
\boldsymbol{\mathcal{H}}_t = [\boldsymbol{\mathcal{H}}_{0,t}]_{:,:,m_{\text{s}} : m_{\text{s}} + M -1}
=\llbracket \boldsymbol{\beta}_t; \mathbf{A}_{1,t}, \mathbf{A}_{2,t}, \mathbf{A}_{3,t}\rrbracket,
\end{split}
\label{app1:trch}
\end{equation}
where $m_{\text{s}}$ denotes the starting index and $\mathbf{A}_{3,t}$ consists of a group of $M$ rows of $\bar{\mathbf{A}}_{3,t}$. Note that the parameters $\{\beta_{r,t}\}_{r=1}^R, \{\tau_{r,t}\}_{r=1}^R, \{\Theta_{1, r, t}\}_{r=1}^R$ and $\{\Theta_{2, r, t}\}_{r=1}^R$ determine $\boldsymbol{\beta}_t$, $\mathbf{A}_{1,t}$, $\mathbf{A}_{2,t}$ and $\mathbf{A}_{3,t}$, respectively. For the tensor channel, the BS only activates one RF chain to transmit a pilot symbol $x_{n_{2}}$ using a beamforming vector $\mathbf{f}_{2, n_{2}}\in\mathbb{C}^{I_{2}\times 1}$. During the transmission of the pilot $x_{n_{2}}$ at the BS, the MS employs $N_{1}$ combining vectors $\mathbf{f}_{1,1},\ldots,\mathbf{f}_{1,N_{1}}\in\mathbb{C}^{I_{1}\times 1}$ to process it. Note that the RF chains of the MS need to be reused when its total number is less than $N_{1}$. After transmitting $N_{2}$ pilots, the received baseband signal tensor can be written as
\begin{equation}
\begin{split}
\boldsymbol{\mathcal{Y}}_t \!= \!\boldsymbol{\mathcal{H}}_t\!\times_1\! \mathbf{F}_{1}^H\!\times_2 \text{diag}(\mathbf{x})\mathbf{F}^T_{2} \!+\! \boldsymbol{\mathcal{N}}_t\!\times_1 \! \mathbf{F}_{1}^H\in\mathbb{C}^{N_{1}\!\times\! N_{2}\!\times\! M},
\end{split}
\label{app1:trsig}
\end{equation}
where $\mathbf{x} = [x_{1},\ldots,x_{N_{2}}]^T$, $\boldsymbol{\mathcal{N}}_t\in\mathbb{C}^{I_{1} \times N_{2}\times M}$ is the noise tensor. $\mathbf{F}_{1}=[\mathbf{f}_{1,1},\ldots,\mathbf{f}_{1,N_{1}}]\in\mathbb{C}^{I_{1}\times N_{1}}$ and $\mathbf{F}_{2}=[\mathbf{f}_{2,1},\ldots,\mathbf{f}_{2,N_{2}}]\in\mathbb{C}^{I_{2}\times N_{2}}$ are the known combiner and beamformer, respectively.
Suppose that $\mathbf{x}=\mathbf{1}_{N_{2}}$ and the rows of $\mathbf{F}_{1}$ and $\mathbf{F}_{2}$ are orthogonal with $N_{1} = I_{1}$ and $N_{2} = I_{2}$. Then, according to (\ref{app1:trch}) and (\ref{app1:trsig}), we can obtain a coarse estimation of $\boldsymbol{\mathcal{H}}$, which is processed as
\begin{equation}
\begin{split}
\bar{\boldsymbol{\mathcal{H}}}_t &= \boldsymbol{\mathcal{Y}}_t\times_1\mathbf{F}_{1}\times_2 \mathbf{F}^*_{2}
\\
&= \boldsymbol{\mathcal{H}}_t \times_1 \mathbf{F}_{1}\mathbf{F}_{1}^H \times_2 \mathbf{F}^*_{2}\mathbf{F}^T_{2} + \boldsymbol{\mathcal{N}}_t \times_1   \mathbf{F}_{1}\mathbf{F}_{1}^H\times_2 \mathbf{F}^*_{2}
\\
&= \boldsymbol{\mathcal{H}}_t + \boldsymbol{\mathcal{N}}_t \times_2 \mathbf{F}^*_{2}
\\
&= \llbracket \boldsymbol{\beta}_t; \mathbf{A}_{1,t}, \mathbf{A}_{2,t}, \mathbf{A}_{3,t}\rrbracket+ \boldsymbol{\mathcal{N}}_t \times_2 \mathbf{F}^*_{2}.
\end{split}
\label{app1:trchno}
\end{equation}
In this way, one can collect plentiful samples $\{\bar{\boldsymbol{\mathcal{H}}}_t\}$ at the receiver. As the least-squares-based coarse estimation can not remove the noise, the obtained a dataset of the noisy channels that characterizes a distribution $p(\bar{\boldsymbol{\mathcal{H}}})$. Fortunately, the proposed DL-CPALS can be trained by only using the noisy tensor $\bar{\boldsymbol{\mathcal{H}}}$ according to Algorithm \ref{alg:dlcpals}. Once the model is trained, in the inference phase, the pre-estimated channel is used as input to the model $\Psi_{\boldsymbol{\theta}}$ in order to obtain favorable initialization factors. Then, the ALS with a small number of iterations is calculated to compute the estimated (denoised) channel. In addition, one can also estimate the channel parameters for each sample $\bar{\boldsymbol{\mathcal{H}}}_t$, such as the spatial AoAs, spatial AoDs and delays from the decomposed factors $\mathbf{A}_{1,t}$, $\mathbf{A}_{2,t}$ and $\mathbf{A}_{3,t}$ by solving the optimization problem (\ref{est_z}).

\section{Simulations}
In this section, we evaluate the tensor CP decomposition and channel estimation performance of the proposed DL-CPALS in the considered mmWave MIMO-OFDM systems.

\subsection{Results on synthetic data}
\label{sec:syn}

We first compare the low-rank CP approximation performance of the proposed DL-CPALS with CPALS with random initializations (Random-CPALS), CPALS with SVD-based initializations (SVD-CPALS) and CPALS with GEVD-based initializations (GEVD-CPALS)\cite{vervliet2016tensorlab}\footnote{https://www.tensorlab.net/}. Consider a noisy low-rank CP tensor distribution $p(\boldsymbol{\mathcal{X}}_{\text{no}})$ with $\boldsymbol{\mathcal{X}}_{\text{no},t} = \boldsymbol{\mathcal{X}}_{t} + \boldsymbol{\mathcal{N}}_{t}\in\mathbb{R}^{6\times 6\times 6}$, where the CP rank of $\boldsymbol{\mathcal{X}}_{t}$ is set as $R=3$ and $\boldsymbol{\mathcal{N}}_{t}$ is Gaussian noise. Based on the CP model of $\boldsymbol{\mathcal{X}}_{t}$ in (\ref{cp1}), the elements of the factors $\{\mathbf{A}_{n,t}\in\mathbb{R}^{6\times 3}\}_{n=1}^3$ and the weight vector $\boldsymbol{\alpha}_{t}\in\mathbb{R}^{3\times 1}$ are all drawn from the uniform distribution $\mathcal{U}(0,1)$. 
The factors $\{\mathbf{A}_{n,t}\in\mathbb{R}^{6\times 3}\}_{n=1}^3$ have full column rank almost surely due to the randomness. According to Theorem \ref{The1}, the constructed tensor is unique if $R<3.5$. Hence $R=3$ satisfies the sufficient condition of uniqueness.
Then, by setting the signal-to-noise-ratio (SNR) to $15$dB, we generate $100000$ noisy tensors randomly of $\{\boldsymbol{\mathcal{X}}_{\text{no},t}\}_{t=1}^{100000}$, where $L_{\text{tr}} = 90000$ samples are used for training and $L_{\text{te}} = 10000$ samples are used for testing. For the DNN model as described in Subsection \ref{sec:DNN}, we set the number of hidden layers to $D=4$, the number of neurons to $Q=512$ and the dropout rate to $0.3$. Thus, the total number of model parameters is $917540$. For the model training as given in Algorithm \ref{alg:dlcpals}, the DL-CPALS uses the Adam optimizer \cite{KingmaB14} with a learning rate $0.00002$ and mini-batch size $L=512$. The number of learning steps is decided by the number of epochs, which is set to $N_{\text{epoch}} = 1000$. As for the setting of the iteration number $K$, we set $K=2$ for training because using a larger $K$ does not lead to significant gains in the experiments but bring more complexity for calculating the gradients. Unless otherwise specified, $K$ is set to $50$ for testing. The performance of the competing CPALS method is evaluated from two aspects: the iterative behavior of the objective value, i.e.,
\begin{equation}
\begin{split}
\frac{1}{L_{\text{te}}}\sum_{t=1}^{L_{\text{te}}}\left\|\boldsymbol{\mathcal{X}}_{\text{no}, t} - \llbracket \mathbf{1}_{R}; \hat{\mathbf{A}}^K_{1,t}, \hat{\mathbf{A}}^K_{2,t}, \hat{\mathbf{A}}^K_{3,t}\rrbracket\right\|^2_F,
\end{split}
\end{equation}
and the iterative behavior of the average NSE (ANSE) \footnote{Note that each true sample $\boldsymbol{\mathcal{X}}_t$ has a single estimation trial $\llbracket \mathbf{1}_{R}; \hat{\mathbf{A}}^K_{1,t}, \hat{\mathbf{A}}^K_{2,t}, \hat{\mathbf{A}}^K_{3,t}\rrbracket$, so we call it ANSE for rigorousness.}, i.e,
\begin{equation}
\begin{split}
\frac{1}{L_{\text{te}}}\sum_{t=1}^{L_{\text{te}}}\frac{\left\|\boldsymbol{\mathcal{X}}_t - \llbracket \mathbf{1}_{R}; \hat{\mathbf{A}}^K_{1,t}, \hat{\mathbf{A}}^K_{2,t}, \hat{\mathbf{A}}^K_{3,t}\rrbracket\right\|^2_F}{\left\|\boldsymbol{\mathcal{X}}_t\right\|^2_F}.
\end{split}
\end{equation}

\begin{figure}[!t]
\centering
\includegraphics[width=0.8\linewidth]{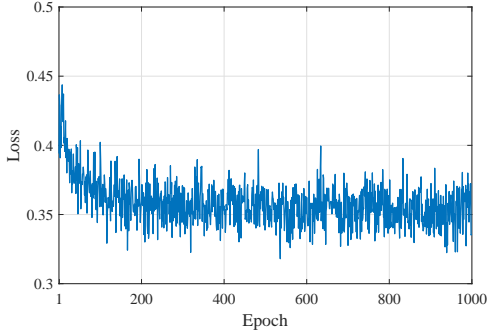}
\caption{The training loss of DL-CPALS.}
\label{fig:loss}
\end{figure}

\par
\begin{figure}[!t]
\centering
\includegraphics[width=1\linewidth]{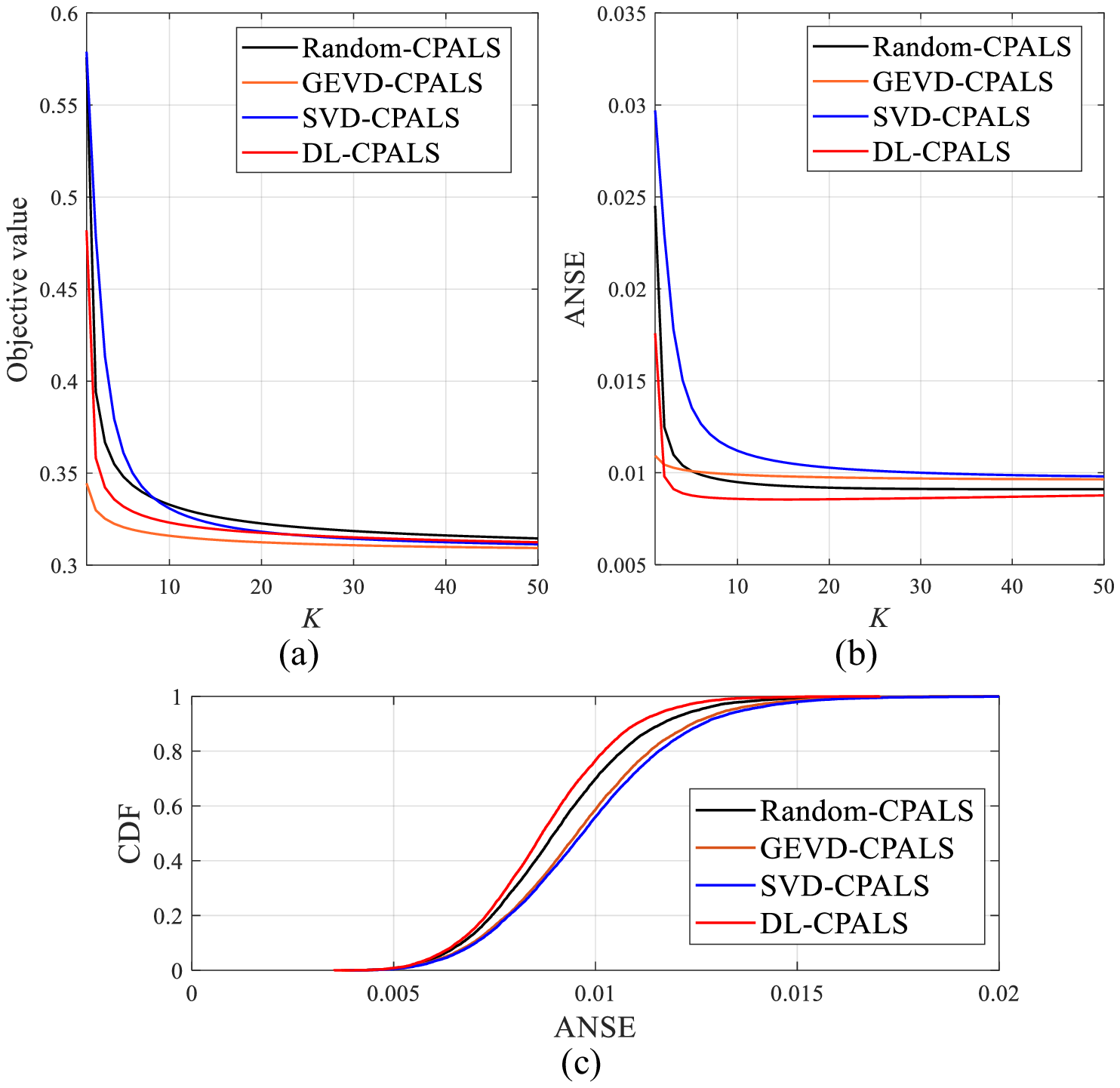}
\caption{The performance comparison of CPALS using different initializations on synthetic data. (a) Iterative behavior of the objective value. (b) Iterative behavior of the ANSE. (c) Curves of the cumulative distribution function (CDF) of the ANSE with $K=50$. }
\label{fig:04}
\end{figure}

In Figure \ref{fig:loss}, we first show the training loss of the proposed DL-CPALS. It can be seen that the loss value generally decreases and tends to converge as the number of epochs increases, which demonstrates that the proposed DL-CPALS is learnable.
In Figure \ref{fig:04}, we plot the performance versus number of iterations of all test samples using different initializations. Note that the elements of the initial factors for the Random-CPALS obey the uniform distribution $\mathcal{U }(0,1)$, which actually means non-negative priors are used.
Figure \ref{fig:04}(a) shows that the convergence of the proposed DL-CPALS is the steepest, because the trained model produces favorable initializations that reduce the objective value quickly. Although the GEVD makes a good initial guess, it fails to consider the updating steps of CPALS in the sequel.
Moreover, the objective value with $K=2$, i.e., the test loss after training is complete, is $0.36$ in Figure \ref{fig:04}(a), which is close to the final training loss in Figure \ref{fig:loss}.
In Figure \ref{fig:04}(b), it can be seen that the proposed method achieves the lowest reconstruction error by only using a small number of iterations (about $K=5$), while the competing Random-CPALS, GEVD-CPALS and SVD-CPALS fail to reach the same accuracy as DL-CPALS even with $K=50$. In order to observe the ANSE distribution of the competing methods on all testing samples, we plot the CDF curves in Figure \ref{fig:04}(c). It can be seen that DL-CPALS has the highest probability of reaching a specific low error.
Based on these results, the proposed DL-CPALS leads to an accurate and fast low-rank CP decomposition.

\subsection{Results of MIMO-OFDM channel estimation}
\label{CEresults}
Consider the ULAs of the BS and the MS are of size $I_{2}=32$ and $I_{1}=8$, respectively. The inter-element spacing $d$ of the ULAs is set to half the carrier wavelength $\lambda$. The sampling rate is set to $f_{\text{b}} = 0.32$GHz. The total number of subcarriers is set to $M_0=128$, out of which a pilot block is selected randomly with $M=4$ subcarriers and $m_{\text{s}}\in\{1,\ldots,125\}$. Suppose that the elements of the noise tensor $\boldsymbol{\mathcal{N}}_t$ are i.i.d. $\mathcal{CN}(0, \sigma^2)$, where $\sigma$ is determined by the SNR.
The AoAs $\vartheta_{r,t}$, AoDs $\theta_{r,t}$, delays $\tau_{r,t}$ in nanoseconds and gains $\beta_{r,t}$ are i.i.d. with distribution $\mathcal{U}(-\frac{\pi}{2},\frac{\pi}{2})$, $\mathcal{U}(-\frac{\pi}{2},\frac{\pi}{2})$, $\mathcal{U}(0,100)$ and $\mathcal{CN}(0, 1)$ for $r=1,\ldots, R$, respectively. We generate $100.000$ samples, where $10.000$ samples are selected randomly for testing and the rest 90.000 samples are used for training.
The number of channel paths is set to $R=4$. 
As the generators are distinct, the Vandermonde factors have full Kruskal rank according to \cite{942615}. Based on Theorem \ref{The1}, $R=4$ satisfies the sufficient condition of uniqueness in this subsection, i.e., $R<5$.
Note that $R$ is known for the proposed method as a priority. One can also determine it as an unknown value by using the information theoretic criteria of minimum description length (MDL) \cite{wax1985detection}.
The fully connected neural network structure has the setting described in Subsection \ref{sec:syn} with $4$ hidden layers and $512$ neurons per layer, where the input layer dimension is $2I_{2}I_{1}M=2048$, and the output layer dimension is $2R(I_{1}+M)=96$. 
Therefore, the total number of parameters is 1886304.
In order to improve the practicality of the model in actual scenarios, note that the DNN is trained by using the channel samples with mixed SNRs which are selected randomly from $\{5, 10, 15, 20, 25\}$dB and tested for each SNR.

\par
For the proposed DL-CPALS method, a two-stage training scheme is adopted. In the first stage, we set the number of iterations $K=0$ (see Algorithm \ref{alg:1}) and train the DNN for 1000 epochs with learning rate $0.0001$, which sets the parameters to a reasonable value range so that the reconstructed CP tensor is close to the input. Then, in the second stage, we set $K=2$ and the learning rate to $0.00002$ for training another 1000 epochs. 
\par
As for the competing methods, in addition to Random-CPALS and SVD-CPALS, the supervised deep-learning-based MIMO-OFDM channel estimation method SFCNN \cite{dong2019deep} is also considered as a baseline, which uses noisy channel samples and noise-free samples to train a convolutional neural network for denoising. The SFCNN is trained for 2000 epochs with learning rate $0.001$. Other settings are the same as in the open-source codes of the paper\footnote{https://github.com/phdong21/CNN4CE}.
As listed in Table \ref{tab:t1}, subspace-type methods estimate the signal subspace and channel parameters by exploiting the Vandermonde structure of the factors, which can also reconstruct an estimated channel using the estimated channel parameters. When applied to the channel estimation of MIMO-OFDM systems, a subspace-type method SCPD (Structured CP decomposition) \cite{9049103} can be obtained, which is also considered as a benchmark.
In addition, we also compare the competing methods with the minimum mean-squared error (MMSE) estimator. According to (\ref{app1:trsig}), the signal model can be written as $\mathbf{y}_t = \mathbf{X}\mathbf{h}_t+\mathbf{n}_t$, where $\mathbf{X} = \mathbf{I}_{M}\otimes \text{diag}(\mathbf{x})\mathbf{F}^T_{2} \otimes \mathbf{F}_{1}^H$, and $\mathbf{y}_t$, $\mathbf{h}_t$ and $\mathbf{n}_t$ are the vectorization forms of $\boldsymbol{\mathcal{Y}}_t$, $\boldsymbol{\mathcal{H}}_t$ and $\boldsymbol{\mathcal{N}}_t\!\times_1 \! \mathbf{F}_{1}^H$, respectively. Then, an MMSE estimation is written as $\hat{\mathbf{h}}_t = \hat{\mathbf{W}}\mathbf{y}_t$, where $\hat{\mathbf{W}}$ is given as
\begin{equation}
\begin{split}
\hat{\mathbf{W}} = &\arg\min_{\mathbf{W}}\mathbb{E} \{\left \|\mathbf{h}_t-\mathbf{W}\mathbf{y}_t\right\|_2^2\}
\\
= & \mathbb{E} \{\mathbf{h}_t\mathbf{h}_t^H\}\left(\mathbb{E} \{\mathbf{h}_t\mathbf{h}_t^H\} + \mathbb{E} \{\mathbf{n}_t\mathbf{n}_t^H\}\right)^{-1}\mathbf{X}^H.
\end{split}
\end{equation}
Note that the covariance matrix of the channel is estimated from all training samples and the covariance matrix of the noise is set as true according to the SNR.

\begin{figure}[!t]
\centering
\includegraphics[width=1\linewidth]{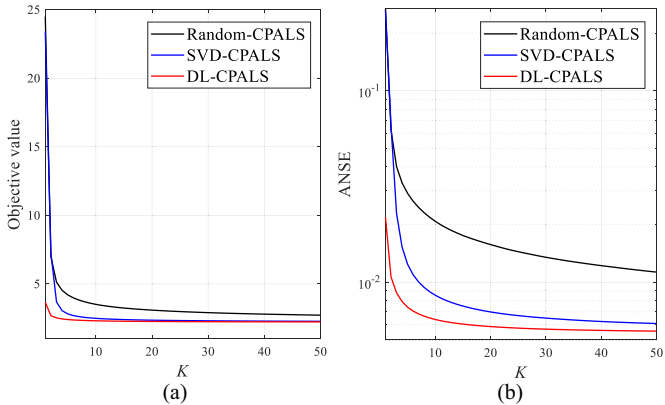}
\caption{The performance comparison of CPALS using different initializations on channel data. (a) Iterative behavior of the objective value. (b) Iterative behavior of the ANSE.}
\label{fig:05}
\end{figure}

\par
We first demonstrate the performance gain of the proposed DL-CPALS for CP decomposition on efficiency and accuracy in Figure \ref{fig:05}, where SNR $=15$dB. Figure \ref{fig:05}(a) shows that the proposed DL-CPALS can generate favorable initializations for the CP decomposition that make the objective function decrease rapidly. 
Note that a better channel estimation performance always corresponds to a lower ANSE but may not to the lower objective value.
Although as $K$ increases, SVD-CPALS can achieve an objective function value similar to that of DL-CPALS, the channel estimation results of the proposed DL-CPALS are more accurate as shown in Figure \ref{fig:05}(b).

\begin{figure}[!t]
\centering
\includegraphics[width=0.85\linewidth]{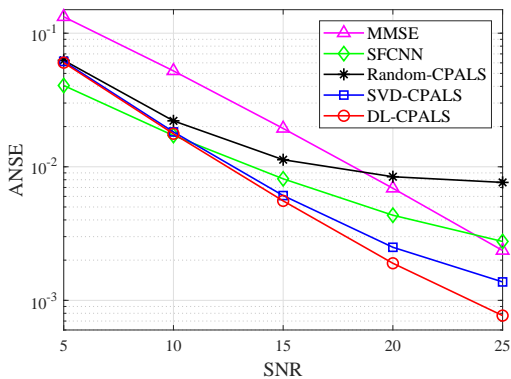}
\caption{The ANSE of the estimated channels for different SNRs.}
\label{fig:06}
\end{figure}

\begin{figure}[!t]
\centering
\includegraphics[width=1\linewidth]{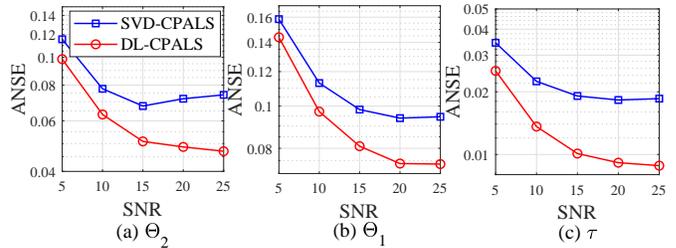}
\caption{ANSE versus SNR of estimated channel parameters including (a) spatial AoD $\Theta_2$, (b) spatial AoA $\Theta_1$ and (c) delay $\tau$}
\label{fig:07}
\end{figure}

\par
Figure \ref{fig:06} shows the comparison of the channel estimation ANSE between the supervised SFCNN, the subspace-type SCPD and the CPALS algorithm with different initializations under different SNRs with $K=50$. 
It can be seen that the performance of SFCNN at SNR $=5$dB is better than that of the SCPD and the CPALS algorithm, because the DL-CPALS is trained only using noisy data and a low SNR may hurt CPALS for solving a CP low-rank approximation, whereas the SFCNN is trained specially for channel denoising by using the paired clean samples and noisy samples. Nevertheless, the performance of the proposed DL-CPALS becomes optimal as the SNR increases. In addition to the denoising effect, the proposed method can also decompose the high-dimensional channel tensor into low-dimensional factors, which can save the channel storage and further reduce the precoding complexity by using the decomposed factors \cite{9443492}.
Although MMSE describes a nearly optimal linear estimation performance, it cannot achieve the accuracy of the DL-CPALS that exploits the channel low-rankness. Moreover, we provide the channel parameter estimation results of SVD-CPALS and DL-CPALS in Figure \ref{fig:07}, where the true parameters and estimated ones are all sorted in descending order due to the ambiguity. Note that the SFCNN cannot perform parameter estimation and the results of random-CPALS are poor, which is therefore not provided in the figure. It can be seen that the parameter estimation error of the proposed DL-CPALS is better than SVD-CPALS in all cases.

\subsection{Complexity analysis}

In this subsection, we discuss the computational complexity of the proposed DL-CPALS in the testing phase. The experiments are all implemented through Python and PyTorch programming on a single computer, which is equipped with an Intel(R) Core(TM) i7-9700K 3.60GHz CPU and 32G memory. During the testing phase, all methods are evaluated on the CPU. We consider SNR $=20$dB and focus on the channel estimation experiments of Subsection \ref{CEresults}. We set the maximum number of iterations to $K=500$.

\begin{table}[t!]
\centering
  \caption{The number of iterations of different CPALSs for achieving the same ANSE of channel estimation} \label{tab:t2}
  \renewcommand\arraystretch{1.15}
\begin{tabular}{cccc}
\toprule
ANSE & Random-CPALS& SVD-CPALS& DL-CPALS\\
\midrule
0.01   & $36$  & $5$  & $2$                  \\
0.004  & $259$ & $15$ & $5$                   \\
0.002  & Fail  & $158$ & $36$                 \\
\bottomrule
\end{tabular}
\end{table}

\begin{table}[t!]
\centering
  \caption{Complexity comparison of different initialization methods of CPALS}
  \label{tab:t3}
  \renewcommand\arraystretch{1.35}
\begin{tabular}{cccc}
\toprule
\multirow{2}{*}{Method} & \multirow{2}{*}{Time Complexity $\mathcal{O}$} & \multicolumn{2}{c}{Kilo Cflops}  \\
                        &          & Init. & Per Iter.                            \\
\midrule
Random                &     \makecell{$KI_{1} I_{2 } M R$\\$+ K(I_{1} I_{2} + I_{1} M + I_{2} M) R^2$}           &  -       & 40   \\
\midrule
SVD                   &  \makecell{$I_{1} I_{2 } M(I_{1}+M) +KI_{1} I_{2 } M R$\\$+ K(I_{1} I_{2} + I_{1} M + I_{2} M) R^2$} &  $\sim$16             & 40            \\
\midrule
DL                    &  \makecell{$DI_{1} I_{2 } M+QD^2 +KI_{1} I_{2 } M R$\\$+ K(I_{1} I_{2} + I_{1} M + I_{2} M) R^2$}&  314             & 40            \\
\bottomrule
\end{tabular}
\end{table}

\par
We report the number of iterations required for CPALS to achieve a certain channel estimation accuracy using different initialization approaches in Table \ref{tab:t2}, where we can see that DL-CPALS greatly saves iterations to guarantee a certain accuracy. For example for the ANSE $=0.002$, the proposed DL-CPALS only takes 36 iterations, while SVD-CPALS takes 158 iterations and Random-CPALS fails to achieve the prescribed accuracy even after 500 iterations.
\par
Compared with Random-CPALS, both SVD-CPALS and DL-CPALS require some complexity for the initializations. In Table \ref{tab:t3}, we first show the computational complexity of these methods. The initializations of SVD-CPALS and DL-CPALS lead to a complexity of $I_{1} I_{2 } M(I_{1}+M)$ and $DI_{1} I_{2 } M+QD^2$, respectively, where $I_{1}$, $I_{2}$, $M$, $Q$ and $D$
are the numbers of receiving antennas, transmitting antennas, training subcarriers, nodes of each hidden layer and hidden layers, respectively.
Although a simple fully connected network is employed to demonstrate the effectiveness of our method, the DNN structure could be more dedicated such as a convolutional network so that its complexity is comparable with the SVD. Moreover, the number of floating point operations (flops) is also provided in Table \ref{tab:t3}. A flop serves as a basic unit of real number computation, which could denote one addition, subtraction, multiplication or division of real floating point numbers. Denote that a complex flop (cflop) serves as a basic unit of complex number computation. Due to that a complex multiplication requires 6 real flops, a cflop is more expensive than a flop. Suppose that $\mathbf{A}\in\mathbb{C}^{m\times n}$, $\mathbf{B}\in\mathbb{C}^{k\times n}$, $\mathbf{C}\in\mathbb{C}^{n\times l}$ and positive definite $\mathbf{R}\in\mathbb{C}^{n\times n}$. According to [54], one can obtain that the numbers of cflops of $\mathbf{A}^H\mathbf{A}$, $\mathbf{A}\mathbf{C}$, $\mathbf{A}\diamond\mathbf{B}$, $\mathbf{A}\odot\mathbf{B}$ and $\mathbf{R}^{-1}$ are $mn^2+mn-\frac{n^2}{2}-\frac{n}{2}$, $2mnl-ml$, $mkn$, $mn$ and , $n^3+n^2+n$, respectively, where the matrix inverse is calculated based on the Cholesky decomposition. As for the SVD, the popular bidiagonalization algorithm produces $\sim \frac{8}{3}n^3$ cflops \cite{trefethen2022numerical}, where $\sim$ means that only the leading term is considered. Thus, we can obtain that each iteration of CPALS takes $40$ kilo cflops. The SVD initialization takes $16$ kilo cflops at least, which is roughly equivalent to one iteration. According to the flops counter THOP\footnote{https://github.com/Lyken17/pytorch-OpCounter} of the DNN model for PyTorch, the DNN model of the proposed DL-CPALS requires 1884160 flops, which is about $314$ kilo cflops. The storage of the DNN is 7.2 megabytes. Although the DL initialization consumes about 8 iterations of cflops, it saves more than 100 iterations of cflops at ANSE of $0.002$ in Table II.
In addition, the average initialization time of a single sample is about $0.6$ milliseconds. Compared with the running time it took for the ALS iterations, e.g., about $22$ milliseconds with $K=50$, the time required for computing the initializations is negligible.

\section{Conclusion}
This paper proposes a deep-learning-aided CP alternating least squares (DL-CPALS) method for efficient tensor CANDECOMP/PARAFAC (CP) decomposition. Specifically, by integrating the model-driven CP decomposition into the data-driven learning process, a neural network model is trained to generate favorable initializations for a fast and accurate CP decomposition. Due to the CP low-rankness constraint exploited in the model, the proposed DL-CPALS only requires noisy samples and does not require paired noise-free samples. The method is applied to the CP low-rank approximation using synthetic data and tensor channel estimation for millimeter wave multiple-input multiple-output orthogonal frequency division multiplexing (mmWave MIMO-OFDM) systems. Experimental results show that the proposed method achieves a fast and accurate CP decomposition compared with the baseline methods.

%
%

\ifCLASSOPTIONcaptionsoff
  \newpage
\fi



%

\bibliographystyle{IEEEtran}
\bibliography{IEEEabrv,mybibfile}

\end{document}